\documentclass{eptcs}
\providecommand{\event}{DCM 2016} 
\usepackage{amsmath,amssymb,amsfonts}
\usepackage{multirow}

\title{Formalising Confluence in PVS}
\author{Mauricio Ayala-Rinc\'on
\institute{Departamentos de Ci\^encia da Computa\c{c}\~ao e Matem\'atica\\  Universidade de Bras\'ilia, Brazil}
\email{ayala@unb.br}}
\def\titlerunning{Formalising Confluence in PVS}
\def\authorrunning{M. Ayala-Rinc\'on}
\begin{document}
\maketitle

\begin{abstract}
  Confluence is a critical property of computational systems which is
  related with determinism and non ambiguity and thus with other
  relevant computational attributes of functional specifications and
  rewriting system as termination and completion.  Several criteria
  have been explored that guarantee confluence and their
  formalisations provide further interesting information.  This work
  discusses topics and presents personal positions and views related
  with the formalisation of confluence properties in the Prototype
  Verification System PVS developed at our research group.
\end{abstract}

\section{Introduction}

Syntactic criteria such as avoiding overlapping of rules as well as
linearity of rules have been used as a discipline of functional
programming which avoids ambiguity. In the context of term rewriting
systems (TRSs for short), well-known results such as Newman's Lemma
\cite{newman}, Rosen's Confluence of Orthogonal term rewriting systems
\cite{Ro1973} as well as the famous Knut-Bendix(-Huet) Critical Pair
Lemma \cite{KnBe70,Huet81} are of great theoretical and practical
relevance.  The first one, guarantees confluence of Noetherian and
locally confluent abstract reduction systems; the second one, assures
confluence of \emph{orthogonal} term rewriting systems, that are
systems that avoid ambiguities generated by overlapping of their rules
and whose rules do not allow repetitions of variables in their
left-hand side (i.e., left-linear); and, the third one provides local
confluence of term rewriting systems whose critical pairs are
joinable.

Formalisations in PVS of these confluence criteria provide valuable
and precise data about the theory of rewriting (cf. \cite{GaAR2010},
\cite{GaAR2008c}, \cite{ROAR2013}).  All mentioned specifications and
formalisations are available either at the local site {\tt
  http://trs.cic.unb.br} or, as part of the NASA PVS libraries, in the
theories for abstract reduction systems {\tt ars} and term rewriting
systems {\tt trs} that belong to the TRS development, at {\tt
  http://shemesh.larc.nasa.gov/fm/ftp/larc/PVS-library/pvslib.html}.

\section{Background}

It is assumed the reader is familiar with rewriting notations and
notions as given in \cite{BaNi98} and \cite{te2003}.

\subsection{Abstract reduction systems}

In the PVS development for TRSs, specifically in the theory {\tt ars},
an abstract reduction system (for short, ARS) is specified as a binary
relation $R$ over an uninterpreted type $T$, {\tt R VAR :
  PRED[[T,T]]}.  This choice facilitates the definition of associated
necessary relations through the use of PVS operations for relations
such as reversal, subset, union, composition and an operator for
iterative applications of compositions. For instance,

\begin{itemize}
\item the inverse of the relation is specified as {\tt converse(R)};
\item the symmetric closure, {\tt SC(R)}, as {\tt union(R,
    converse(R)};
\item the reflexive closure, {\tt RC(R)}, as {\tt union(R, =)};
\item the reflexive transitive closure, {\tt RTC(R)}, as {\tt
    IUnion(LAMBDA n: iterate(R, n))};
\item the equivalence closure, {\tt EC(R)}, as {\tt RTC(SC(R))} etc.
\end{itemize}
     
Properties of ARSs are then specified in a very natural manner from
these relational basis. For instance, using PVS properties for
relations such as well-foundedness, the property of noetherianity,
{\tt noetherian?(R)} is specified as the predicate {\tt
  well\_founded?(converse(R))}.  Also, the property of confluence,
{\tt confluence?(R)}, is specified as

{\tt FORALL x, y, z: RTC(R)(x,y) \& RTC(R)(x,z) => joinable?(R)(y,z)}

where the predicate {\tt joinable?} is specified as

{\tt joinable?(R)(x,y): bool = EXISTS z: RTC(R)(x,z) \& RTC(R)(y, z)}.

More synthetic specifications might be possible; for instance, the
elegant \emph{set-theoretical} definition of confluence, written in
the usual rewriting notation as $(\mbox{}^*\!\leftarrow \circ
\rightarrow^*) \subseteq (\rightarrow^* \circ\;
\mbox{}^*\!\leftarrow)$ can be specified straightforwardly as {\tt
  subset?(RTC(converse(R)) o RTC(R), RTC(R) o RTC(converse(R)))},
using relation composition, {\tt o}, and the subset predicate, {\tt
  subset?}.

ARS results were formalised using standard proof techniques as
noetherian induction. Among other results on confluence of ARSs, a
description of the formalisation of Newman Lemma, specified below, is
available in \cite{GaAR2008c}

{\tt Newman: LEMMA noetherian?(R) => (confluent?(R) <=>
  locally\_confluent?(R))}

\subsection{Term Rewriting Systems}

Terms are specified as a data type built from variables over a
nonempty uninterpreted type and a signature of function symbols with
their respective arities. The arguments of a functional term, headed
by a function symbol of the signature, are specified as a finite
sequence of terms, of length equal to the arity of the function
symbol.  Positions of a term {\tt t}, written {\tt posOF(t)}, are
finite sequences of naturals specified recursively as in the standard
way in the theory of rewriting.  Thus, the necessary operations on
positions as their concatenation resumes to concatenation of finite
sequences of naturals and so, predicates such as disjunct or parallel
positions, given by the predicate {\tt parallel?} or for short {\tt
  ||}, are specified as {\tt (NOT p <= q) \& (NOT q <= p)}, where {\tt
  <=} is the sequence prefix relation built as {\tt <=(p, q): bool =
  (EXISTS (p1: position): q = p o p1)}.

Using these types for terms and positions, it is possible to build the
required algebraic properties for terms, positions, subterms and
replacement of terms. Namely, the subterm of a term {\tt t} at a valid
position {\tt p}, usually written as $t_p$, is specified recursively
navigating through the structure of the term according to the naturals
in the position sequence and the arguments of the functional terms
inside {\tt t}: {\tt stOF(t: term, (p: positions?(t))}. Also, the
replacement of the subterm at a valid position {\tt p} of a term {\tt
  s} by another term {\tt t} is built recursively: {\tt replaceTerm(s:
  term, t: term, (p: positions?(s)))}, that will be abbreviated as
{\tt s[t]\_p}.  These design decisions give rise to algebraic
properties that are easily formalised by inductive proofs on these
data structures.  Among other properties, one has formalisations for:

\begin{itemize}
\item Preservation of positions after replacement of subterms either
  positions of replacement:

  {\tt posOF(s)(p) => posOF(s[t]\_p)(p)}

  or parallel positions to the position of replacement:

  {\tt posOF(s)(p) \& posOF(s)(q) \& p||q => posOF(s[t]\_p)(q)}

\item Extension of possible new valid positions at the position of
  replacement after a replacement (where below {\tt o} stands for the
  concatenation of sequences):

  {\tt posOF(t)(q) \& posOF(s)(p) => posOF(s[t]\_p)(p o q)}

\item Preservation of replaced terms:

  {\tt posOF(s)(p) \& posOF(t)(q) => stOF(s[t]\_p,p o q) = stOF(t, q)}

\item Associativity of replacement:

  {\tt posOF(s)(p) \& posOF(t)(q) => (s[t]\_p)[r]\_(p o q) =
    s[t[r]\_q]\_p }

\item Commutativity of replacement at parallel positions:

  {\tt posOF(s)(p) \& posOF(s)(q) \& p||q => (s[t]\_p)[r]\_q =
    (s[r]\_q)[t]\_p}

\end{itemize}
  
Rewriting rules are specified as pairs of terms $(l, r)$ satisfying
the usual conditions on rules, that is the left-hand side (lhs) cannot
be a variable and the variables occurring in the right-hand side (rhs)
should belong to the lhs of the rule:
\[\mbox{\tt rewrite\_rule?(l,r): bool = (NOT vars?(l)) \& subset?(Vars(r),
  Vars(l))}\]

After that, it is possible to define the type of rewriting rules as

\[\mbox{\tt rewrite\_rule : TYPE rewrite\_rule?} \]

\noindent and then, a TRS is given as a set of rewriting rules {\tt set[rewrite\_rule]}.

Substitutions are built as objects of type {\tt [V -> term]}, where
{\tt V} is a countably infinite set of variables and such that their
domain is finite, that is {\tt Sub?(sig): bool =
  is\_finite(Dom(sig))}, where {\tt Dom(sig): set[(V)] = \{x: (V) |
  sig(x) /= x\} }. The type of substitutions is given as {\tt Sub:
  TYPE = (Sub?)}.  From this point, renaming, variants, composition of
substitutions and homomorphic extensions of substitutions, {\tt
  ext(sigma)}, are easily built as well as a series of necessary
substitution properties formalised.

With these elements of formal design it is possible to define the
reduction relation from a set of rewriting rules say {\tt E}:

\begin{verbatim}
 reduction?(E)(s,t): bool =  
     EXISTS ( (e | member(e, E)), sigma, (p: positions?(s))) : 
                      stOF(s, p) = ext(sigma)(lhs(e)) & 
                                    t = s[ext(sigma)(rhs(e))]_p 
\end{verbatim}

Immediately, it is possible to prove that this relation is
\emph{closed under substitutions} and \emph{compatible with contexts}.

After that, a predicate for critical pairs of a TRS {\tt E} is built,
{\tt CP?(E)}, and then the most famous result on confluence of TRSs,
that is the Critical Pair Lemma is formalised as described in
\cite{GaAR2010}.

\begin{verbatim}
CP: THEOREM FORALL E: locally_confluent?(reduction?(E)) <=> 
      (FORALL s, t) : CP?(E)(s, t) => joinable?(reduction?(E))(s,t)
\end{verbatim}

Since the reduction relation built from a set of rewriting rules
inherits by parameterisation, all properties of ARSs in the {\tt ars}
development, it is possible to apply Newman's Lemma in order to
formally infer confluence of a terminating TRS all whose critical
pairs are joinable.

The design decisions taken in the specification of ARSs and TRSs were
satisfactory to accomplish one of our main objectives in this
formalisation, that is indeed maintaining formal proofs as close as
possible to the analytical proofs presented in textbooks. In fact,
diagrammatic didactical artefacts (used in papers and textbook)
representing rewriting properties used in the proofs as commutation
diagrams for confluence, local-confluence etc, and those ones used for
representing peaks valleys and overlap situations in the analysis of
the Critical Pair criterion can be also conducted when reasoning about
our formalisations.

In particular, the formalisation of the Critical Pair Lemma follows
the textbook proof organisation which is based on the analysis of the
possible peaks when trying to obtain local confluence. These peaks can
be originated from simultaneous reductions at parallel positions,
which are trivially joinable, or from reductions at nested positions,
which give rise to either \emph{critical} or \emph{non-critical
  overlaps}.

A peak, from a critical overlap can be easily verified to join, by
proving that it corresponds to an instance of a critical pair and
using the assumption that critical pairs are joinable.

The non-critical overlaps are the interesting ones. A such peak is
originated by application of rules $l\rightarrow r$ and $g\rightarrow
d$ with substitution $\sigma$, assuming these rules have not common
variables which is possible by renaming of rules.  Supposing $l\sigma$
occurs in the dominating position of the overlap, one can focus on the
analysis of the \emph{joinability} of the peak $r\sigma \leftarrow
l\sigma \rightarrow l\sigma[d\sigma]_{p\circ q}$, where $p$ is a
variable position in $l$, say $l_p = x$, and $q$ the position in
$x\sigma$ in which $g\sigma$ occurs.

Thus, all that is solved by the careful construction of a new
substitution, $\sigma'$ such that it modifies $\sigma$ mapping $x
\mapsto x\sigma[d\sigma]_q$ and maintains the images of all other
variables in the domain of $\sigma$ as those mapped by $\sigma$.

In the sequel, by a like \emph{``uniform reduction sequence''} one has
that $r\sigma \rightarrow^* r\sigma'$ and $l\sigma[d\sigma]_{p\circ
  q}\rightarrow^* l\sigma'$; the former is done as $r\sigma
\rightarrow r\sigma[d\sigma]_{q_1\circ q}\rightarrow\cdots\rightarrow
r\sigma[d\sigma]_{q_1\circ q}\cdots[d\sigma]_{q_n\circ q} = r\sigma'$,
where $\{q_1,\ldots,q_n\}$ is the set of positions of $r$ in which $x$
occurs, and the latter is done as $l\sigma[d\sigma]_{p\circ
  q}\rightarrow l\sigma[d\sigma]_{p\circ q}[d\sigma]_{p_1\circ
  q}\rightarrow \cdots\rightarrow l\sigma[d\sigma]_{p\circ
  q}[d\sigma]_{p_1\circ q}\cdots[d\sigma]_{p_m\circ q} = l\sigma'$,
where $\{p\}\cup \{p_1,\ldots,p_n\}$ is the set of positions of $l$ in
which $x$ occurs.

So joinability is concluded by the application of rule $l\rightarrow
r$ with substitution $\sigma'$.  See for instance the proof in
Chapters 6 or 2 of respectively \cite{BaNi98} or \cite{te2003}.

The formalisation, under the design choices previously mentioned,
requires the construction of elements that guarantee specialised
properties, such as the instantiation of a critical pair, built from
the rewriting rules, that corresponds to a critical overlap as well as
the substitution $\sigma'$ for a non-critical overlap. For the latter,
it is necessary an inductive proof (using auxiliary lemmas) on the
cardinality of the sets of positions $\{q_1, \ldots, q_n\}$ and
$\{p_1,\ldots,p_m\}$ for concluding that $l\sigma$ and $r\sigma$
rewrite into $l\sigma'$ and $r\sigma'$, respectively.

\section{Formalising the algebra of parallel rewriting and
  orthogonality}

Rosen's confluence of orthogonal TRSs \cite{Ro1973} is a challenging
formalisation.  The classical proof is based on the Parallel Moves
lemma: essentially, what is necessary is to prove that under the
hypothesis of orthogonality, the associated parallel reduction
relation holds the diamond property.

Intuitively, the analytical proof requires only the comprehension of
properties for the notion of the parallel reduction relation, but the
intuition of parallel rewriting is usually explained through the like
``uniform reduction'' as in the analysis of the Critical Pair
criterion, which in fact refers to sequential rewriting. So, any
formalisation following the classical approach would require an
explicit construction of such that parallel relation as well as the
specialised description and formalisation of its specific algebraic
properties.

The notion of parallel reduction depends on sets of triplets of valid
positions, rules and substitutions, positions in which sequential
replacements are simultaneously applied according to the instantiation
of the rules with the associated substitutions.  Because of this
dependence, two design approaches are possible: either using sets of
triplets of positions, rules and substitutions or sets of (finite) and
coordinated sequences of positions $\Pi$, rules $\Gamma$ and
substitutions $\Sigma$. We opted by the last design alternative since
we believe it is closer to implementations in programming languages
and also because PVS offers libraries with translations (and their
necessary formalised properties) between data structures such as sets,
lists and finite sequences.

The parallel rewriting reduction relation built from a set of
rewriting rules $E$, that in classical notation is written as $s
\rightrightarrows t$, is specified as the relation {\tt
  parallel\_reduction?(E)(s,t)} below, using a parallel replacement
operator, {\tt replace\_par\_pos}, that is recursively specified from
the sequential {\tt replaceTerm} operator, and through an auxiliary
relation {\tt parallel\_reduction\_fix?(E)}.

\begin{verbatim}
 parallel_reduction_fix?(E)(s,t, (fsp: SPP(s))): bool =  
    EXISTS ((fse | member(fse, E)), fss) :
           fsp`length = fse`length AND fsp`length = fss`length
           AND subtermsOF(s,fsp) = sigma_lhs(fss, fse)
           AND t = replace_par_pos(s, fsp, sigma_rhs(fss, fse))

 parallel_reduction?(E)(s,t): bool =  
    EXISTS (fsp: SPP(s)): parallel_reduction_fix?(E)(s,t,fsp)
\end{verbatim}

{\tt fsp}, {\tt fse} and {\tt fss} correspond to the sequences $\Pi =
[p_1, \ldots, p_n]$, $\Gamma = [(l_1,r_1), \ldots, (l_n,r_n)]$ and
$\Sigma = [\sigma_1, \ldots, \sigma_n]$ of positions, rewriting rules
and substitutions used in the parallel rewriting.  {\tt fsp} is a
sequence of parallel positions of $s$ obtained by its type dependency,
that is {\tt SSP(s)}. {\tt fse} is a sequence of equations in the
rewriting system given by {\tt member(fse, E)}, and {\tt fss} is a
sequence of substitutions. The required coordination of triplets of
associated positions, rules and substitutions is directly obtained by
using the corresponding indexation in the respective sequences, that
is the same index. The condition {\tt subtermsOF(s,fsp) =
  sigma\_lhs(fss, fse)} equals the condition that for all valid index
$i$ of these sequences, $s_{p_i} = l_i\sigma_i$ and the condition {\tt
  t= replace\_par\_pos(s, fsp, sigma\_rhs(fss, fse))} equals $t$ to
the desired parallel contractum, that is $s$ replacing the
$l_i\sigma_i$'s by the $r_i\sigma_i$'s.

In a parallel peak, say $t \leftleftarrows s \rightrightarrows u$,
using positions, equations and substitutions say $(\Pi_k, \Gamma_k,
\Sigma_k)$ for $k=1,2$, as in the case of the Critical Pair Lemma, the
interesting cases are those of non-critical overlaps.  Without loss of
generality, at some position $q\in \Pi_1$ one has all positions
$p_1,...,p_n$ in $\Pi_2$ below $q$.  On the one side, $s_q = l\sigma$
and $t_q = r\sigma$, where $(l,r)$ is the rewriting rule in $\Gamma_1$
and $\sigma$ the substitution in $\Gamma_1$ associated with the
position $q$ in $\Pi_1$.  On the other side, one has $s_q
\rightrightarrows u_q$ by parallel reduction at positions $p_1, \ldots
p_n$ below $q$ accordingly to the same equations and substitutions in
the triplet $(\Pi_2, \Gamma_2, \Sigma_2)$.  Let $(\Pi', \Gamma',
\Sigma')$ denote the triplet of this last parallel reduction step,
then the parallel peak $r\sigma \leftleftarrows l\sigma
\rightrightarrows u_q$ is an instance of the Parallel Moves Lemma. See
details in Chapter 4.3 of \cite{te2003} or 6.4 of \cite{BaNi98}, for
instance.

Since in such a parallel peak $r\sigma \leftleftarrows l\sigma
\rightrightarrows t$ all overlaps are non critical, one should prove
that parallel reducing each $\sigma$-instance of a variable in $l$,
say $x$ at position $p$ one has $l\sigma_p = x\sigma \rightrightarrows
x\sigma' = t_p$, where the substitution $\sigma'$ is built by reducing
in parallel all occurrences of $\sigma$-instantiated variables in
$l\sigma$ uniformly, which is possible by left-linearity assumption.
Thus, $r\sigma \rightrightarrows r\sigma'$ and $t\rightrightarrows
l\sigma'$, that allows concluding the joinability of the peak.

Despite in the theory the adaptation of sequential properties for
parallel replacement might be intuitively clear, in the PVS
development the necessary specialised algebraic properties should be
formalised. Let {\tt s[T]\_$\Pi$} denote the parallel replacement of
terms in the sequence of terms $T$ at valid parallel positions
$\Pi$. A few of those properties are included below.

\begin{itemize}
\item Preservation of positions after replacement of subterms either
  positions of replacement:

  {\tt posOF(s)($\Pi$) => posOF(s[T]\_$\Pi$)($\Pi$)}

  or parallel positions to the position of replacement:

  {\tt posOF(s)($\Pi$) \& posOF(s)($\Pi'$) \& $\Pi' || \Pi$ =>
    posOF(s[t]\_p)($\Pi'$)}

\item Invariance under composition of parallel replacement at parallel
  sequences of positions:

  {\tt posOF(s)($\Pi_1$) \& posOF(s)($\Pi_2$) \& $\Pi_1 || \Pi_2$ => }

  \hfill {\tt (s[T\_1]\_$\Pi_1$)[T\_2]\_$\Pi_2$ = (s[T\_1 o
    T\_2]\_$\Pi_1\circ\Pi_2$}

\item Commutativity of parallel replacement at parallel sequences of
  positions:

  {\tt posOF(s)($\Pi_1$) \& posOF(s)($\Pi_2$) \& $\Pi_1 || \Pi_2$ => }

  \hfill {\tt (s[T\_1]\_$\Pi_1$)[T\_2]\_$\Pi_2$ =
    (s[T\_2]\_$\Pi_2$)[T\_1]\_$\Pi_1$}

\end{itemize}
    
The formalisation of confluence of orthogonal TRS proceeds by
inductive proof techniques taking care of the specificities of the
algebra of parallel positions, replacement and rewriting.

\section{Conclusions and future work}

The development of these formalisations on confluence of TRSs brought
out several lessons.

From the theoretical point of view, the main observation is that
despite the intuitive notion of "uniform reduction", that is used to
provide intuition about the proof of the Parallel Moves Lemma, induces
to believe that the extension is obvious, a specialised development of
the theory of parallel reduction and its algebraic properties is
necessary. And extending sequential to parallel rewriting
formalisations is not trivial.  A preliminary thoughtful analysis
would be always necessary in order to estimate accurately the real
complexity and the necessary effort of formalisations in any context,
mostly when the proposed development appears to be a simple extension
of those yet available.  The second lesson has to do with the
investment of enough time for fine tuning design decisions since they
influence the effort required in proofs.

The main lesson is that through this kind of exercise, our
comprehension of the theory becomes more refined, providing a better
support for the formal analysis of further related developments, as
well as a more realistic insight about the possible adaptation or
reuse of previous specifications and proofs.  Developments in progress
include use of such techniques in other contexts as the nominal syntax
approach of rewriting (cf.  \cite{ARFeGaRO2015} \cite{ROFeAR2015}).

%\bibliographystyle{eptcs}
%\bibliography{../../../bibliography/bibtra2}

\end{document}